\documentclass[aps,prl,twocolumn]{revtex4}
\usepackage{graphicx}
\usepackage{color}

\begin{document}
\title{Dark Excitons in Carbon Nanotubes}

\author{Eric Chang, Deborah Prezzi, Alice Ruini, and Elisa Molinari}
\affiliation{INFM-CNR National Center on nanoStructures and bioSystems
at Surfaces (S$^3$) and Dipartimento di Fisica, \\
Universit\`{a}
di Modena e Reggio Emilia, Via Campi 213/A, 41100 Modena, Italy \\}

\begin{abstract}
We investigate the lowest many-body
excited states in carbon nanotubes by means of {\it ab initio} calculations. 
On the basis of these calculations and an additional
theoretical analysis of the excitons, we demonstrate that
the splitting between the dark and bright spin-singlet
states is due to the direct Coulomb electron-hole interaction,
that the dark exciton always has 
gerade symmetry, and that it is always lower in energy relative
to the bright exciton.
Applying these studies to 0.4-0.8-nm-diameter tubes, we find a 
gerade-ungerade splitting of several tens of an meV, and a 
singlet-triplet splitting of comparable value.
Our symmetry analysis of the origin of the singlet splitting indicates 
possible startegies to increase the emission efficiency of nanotubes.
\end{abstract}
\maketitle

In recent years, the luminescence properties of carbon nanotubes
have sustained intense theoretical~\cite{pere+04prl,chan+04prl,spat+04prl} 
and experimental research~\cite{bach+02sci,mise+03sci,lefe+03prl,
hage+04apa,wang+04prl,chou+05prl}. 
Of particular interest is the study of the emission efficiency and the radiative
lifetimes, in view of applications in the field
of opto-electronics and photonics~\cite{mise+03sci,frei+03nl,frei+04nl,chen+05sci,pere+05nl}. 

Recent studies on excited-state dynamics~\cite{hage+04apa,wang+04prl,shen+05prb}, 
combined with the experimentally observed low quantum efficiency, 
indicate that exciton relaxation is dominated by non-radiative processes, which 
dramatically quench emission spectra.
The presence of optically forbidden 
states, i.e., {\it dark} states, below the optical gap, 
to which the system can decay non-radiatively, is believed to be responsible for
the luminescence quenching~\cite{zhao-mazu04prl, pere+04prl,pere+05nl}. 
Their presence is estimated to increase the effective radiative lifetime 
of about five times~\cite{spat+05prl}.
However, evidence of dark states has been presented only recently 
in experiments involving
large magnetic fields~\cite{zari+06prl}, and very little is known 
about this subject.

Thus far, no theoretical mechanism has been proposed explaining the
reason why there are dark states lying {\it below} the first bright one nor
if this is a general rule for all kinds of tubes. 

In this work, we carry out {\it ab initio} calculations of
excitonic states for several tubes of sizes and dimensions of 
interest in the experimental physics community. On the basis of our 
calculations, we show that only two pairs of bands are involved
in
 the description of the lowest excitons. This
fact, as we shall see later, implies that
there is always a dark singlet state below the lowest 
bright one. Finally, our symmetry analysis allows us to propose alternative 
ways to increase the quantum efficiency of nanotubes through 
symmetry-breaking mechanisms.

\begin{figure}
\includegraphics[clip,width=0.85\columnwidth]{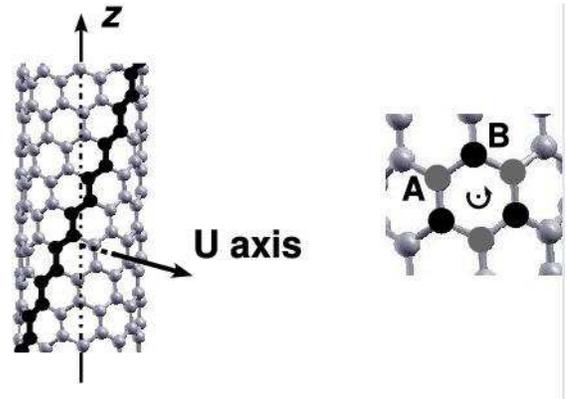}
\caption{\label{symm} 
The two  principal symmetries of carbon nanotubes: the roto-translation,
$RT$, with axis $z$, and the $\pi$-rotation about the axis $U$, $R_U$.
{\it Left panel:} $RT$ is shown by the dark atoms.
The axis $U$ pierces the center of an hexagon
and is perpendicular to the screw axis $z$. {\it Right panel:}
The two inequivalent atoms $A$ and $B$ of each hexagon 
are shown in greater detail. 
When $R_U$ is applied to the atomic positions, $A$ goes into $B$ and viceversa.}
\end{figure}

The excited states are calculated using a many-body theoretical approach
which involves both self-energy corrections in the treatment of single-particle
energies and 
effects arising from the electron-hole 
interaction.
The former is treated within the 
GW approximation~\cite{hedi65pr,hybe-loui86prb}  
while the latter is
obtained from the diagonalization of the 
Bethe-Salpeter Hamiltonian~\cite{albr+98prl,bene+98prl,rohl-loui98prl1,rohl-loui98prl}. 

\begin{figure}
\includegraphics[clip,width=0.68\columnwidth,angle=-90]{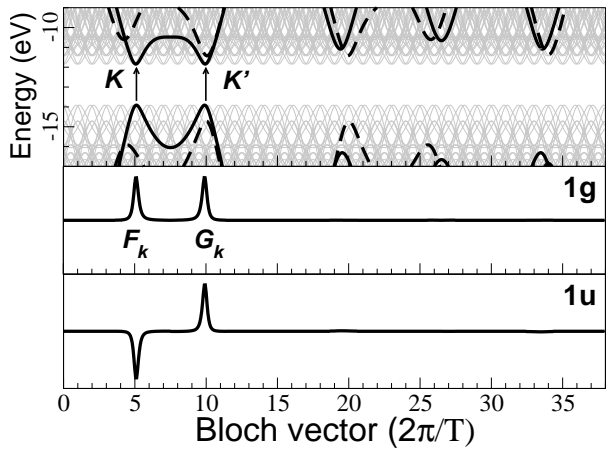}
\caption{\label{band} A typical band structure (upper panel) with the 
wavefunctions for the two lowest excitons (two lower panels) 
in the extended zone scheme. Both wavefunctions are localized
around $K$ and $K'$ and manifest even and odd parity with
respect to the $R_U$ symmetry. The diagrams shown are for the (6,4) tube.
} 
\end{figure}

In our calculations, we take advantage of all the crystal symmetries
of the tube (see Fig.~\ref{symm}), which give us a better understanding of the band structure
as well as of the excited states~\cite{chan+04prl,chan+05prb,maul+05prb}. 
A typical band structure of the tubes, shown in the upper panel
of Fig.~\ref{band} in the extended zone scheme, has a minimum gap
at the points labelled $K$ and $K'$. 

In all the tubes studied, for an energy window from 0 to 5~eV, which is the range of interest for optical absorption experiments, the only transitions
that come into play for the lowest singlet states
are those between the valence band maximum and the conduction band
minimum for $k$-points near $K$ and $K'$.
In fact, for the  
singlet states $1u$ and $1g$, where ``1'' refers to the lowest bound excitonic state and $u$ ($g$) refers to the parity with respect to 
$\pi$-rotations about the axis $U$ (Fig.~\ref{symm}), most of the
weight ($\approx 99\% $) is concentrated near the gap in $K$ and $K'$.
It can be clearly seen in Fig.~\ref{band}, where the probability amplitude of the transition is 
plotted as a function of $k$ in the extended zone scheme. 
In addition, we find that the dark states
have the same excitonic composition, as far as the probability amplitude is
concerned, since the bright states  differ from them only in the phase 
information (lowest panel in Fig.~\ref{band}). 
This behaviour suggests the possibility of describing 
the lowest excitonic states with a graphite-derived, 
four-band model.

\begin{figure}
\vspace{.3cm}
\includegraphics[clip,width=0.68\columnwidth,angle=-90]{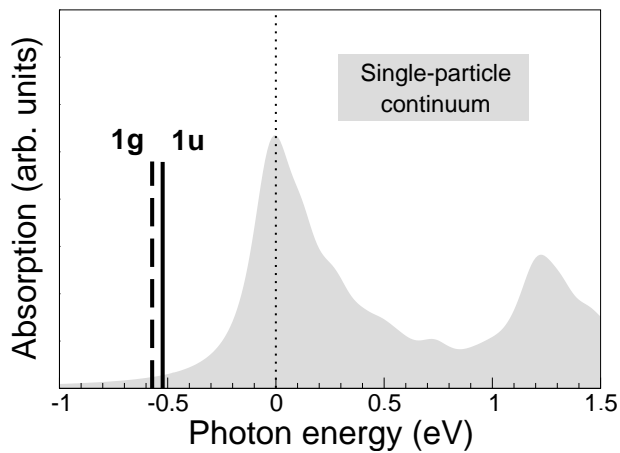}
\caption{\label{spectrum} The lowest singlet states in the absorption 
spectrum of the (6,4) tube.
The shaded line is the single-particle absorption, ignoring excitonic
effects. The solid and dotted vertical lines indicate the positions
of the two lowest excitons ($1g$ and $1u$), where the lower of the two
is dipole forbidden.  
}
\end{figure}
Let us call the excitonic wavefunction concentrated near $K$,
$F_k$, and the corresponding wavefunction at $K'$,
$G_k$.  
Let the indices $i$ and $j$ denote the space of possible
optical transitions $vk \rightarrow ck$ and $H_{ij}$, the
Bethe-Salpeter (BS) Hamiltonian in this space of transitions.
$F_k$ and $G_k$ can be written alternatively in the
basis of the transitions as $F_i$ and $G_i$.

$H_{ij}$ can be decomposed in three parts,
\[
H_{ij} = D_{ij} + 2 X_{ij} - W_{ij} \equiv D_{ij} + I_{ij}.
\]
Here $D_{ij}$ is the diagonal term, $X_{ij}$ is the exchange term, and
 $W_{ij}$ is the direct Coulomb term. $I_{ij} = 2 X_{ij}-W_{ij}$ is
the total electron-hole interaction term.

We  define the parameters $E = E_{gap} - E_{binding}$ and $\Delta$, as
\begin{equation}
E = \sum_{ij} F_i^* H_{ij} F_j = \sum_{ij} G_i^* H_{ij} G_j 
\label{energy}
\end{equation}
and
\begin{equation}
\Delta = \sum_{ij} F_i^* H_{ij} G_j = \sum_{ij} F_i^* I_{ij} G_j .
\label{delta}
\end{equation}
We can obtain excitonic states diagonalizing the matrix
\[ M = \left(
 \begin{array}{cc}
  E & \Delta \\
 \Delta^* & E 
\end{array}
\right).
\]

It has two eigenvectors, ${\bf v}_\pm$, given by
\[
 {\bf v}_{\pm} = {1 \over {\sqrt 2}}\left(
 \begin{array}{c}
  1 \\
 { {\pm \Delta^*} / {|\Delta|}} 
\end{array}
\right)
\]
with eigenvalues, $\lambda_\pm = E \pm |\Delta|.$

We calculate $\lambda_\pm$ using our {\it ab initio} BS 
hamiltonian and model wavefunctions, which take into account
only contributions from transitions at k-points very close to 
 $K$ and $K'$. 
The splitting 
$|\Delta| $ is found with 
an error of no more than $1\%$ with respect to the {\it ab initio} calculation
of the singlet splitting $E_{1u}^S-E_{1g}^S$. We can therefore safely 
conclude that no bands other than the four graphite-derived ones come 
into play for the lowest singlet exciton states and the model 
assumption is correct.

We have yet to demonstrate which of these two states, ${\bf v}_\pm$, is bright. 
In order to simplify the following discussion, 
note that 
the direct term $W$ is large compared
to the exchange term $X$, which  will therefore be neglected in 
what follows with an error less than 1~$\%$. 

We define the real-space excitonic wavefunction 
\[
\Psi_K({\bf r},{\bf r'}) = \sum_k 
F_{k}\phi^*_{c  k} ({\bf r})
     \phi_{v k} ({\bf r'}),  
\]
where $\phi^*_{c  k} ({\bf r})$ and $\phi_{v k} ({\bf r'})$ are  
the single-particle conduction and valence wavefunctions. 

Without loss of generality, we examine only the term with $k=K$, therefore
\[
\Psi_K({\bf r},{\bf r'}) \approx \phi^*_{ck} ({\bf r})
                                          \phi_{vk} ({\bf r'}).
\] 
We exclude the modulation
factor  $F_{k}$ for simplicity, since the argument below would still be valid.
Since the symmetry $R_U$ depicted in Fig.~\ref{symm} transforms
$K$ in $K'$, and consequently, $F_k$ in $G_k$, then Eq.~\ref{delta}
can be written as
\[
\Delta = \int d^3 r \int d^3 r' \Psi_K^*({\bf r},{\bf r'}) W({\bf r},{\bf r'})
\Psi_K({\bf R_U} {\bf r},{\bf R_U}{\bf r'}).
\label{delta:R_U}
\]

If we assume that $\phi_{cK}$ and $\phi_{vK}$ are $\pi$-like, i.e.,
are symmetric ($\pi$) and antisymmetric ($\pi^*$) LCAO combinations of 
basis functions centered on the two inequivalent atomic sites $A$ and $B$ (which
we denote as $\chi_A({\bf r}) $ and $\chi_B({\bf r}) $),
then
$\phi_{cK}({\bf r}) = {1 \over \sqrt{2}} [\chi_A({\bf r}) - e^{i\delta} \chi_B({\bf r}) ]$ and  
$\phi_{vK}({\bf r}) = {1 \over \sqrt{2}} [\chi_A({\bf r}) + e^{i\delta} \chi_B({\bf r}) ]$.

From the relations $\chi_A({\bf R_U}{\bf r}) = \chi_B^*({\bf r})$ and
$\chi_B({\bf R_U}{\bf r}) = \chi_A^*({\bf r}) $, we have
\[
  \Psi_K({\bf R_U} {\bf r},{\bf R_U}{\bf r'}) = -
 \Psi_K^*( {\bf r},{\bf r'}).
\]
Therefore $\Delta$ can be written as
\[
\Delta = - \int \int d{\bf r} d{\bf r'} [\Psi_K({\bf r},{\bf r'})]^2 W({\bf r}, {\bf r'}).
\]
If we define the functions 
$f({\bf r})  =  {1 \over 2} \lbrace [\chi_A({\bf r})]^2 + e^{2i\delta} [\chi_B({\bf r})]^2  \rbrace$
and
$g({\bf r}) = e^{i\delta} \chi_A({\bf r}) \chi_B({\bf r})$,
then $\Delta$ can be written as
\[
  \Delta = \Delta_{AA} + \Delta_{AB},
\]
with
\begin{eqnarray*}
\Delta_{AA} & = & -\int d^3r d^3r' [f({\bf r}) ]^* W({\bf r},{\bf r'}) f({\bf r'}),  \\
\Delta_{AB} & = &\int d^3r d^3r' [g({\bf r}) ]^* W({\bf r},{\bf r'}) g({\bf r'})\\
\end{eqnarray*}
$\Delta_{AA}$ is strictly positive and $\Delta_{AB}$ is strictly negative.
Let the integrated charge of the product of like 
basis functions be defined as 
$Q_{AA} = \int d^3r [\chi_A({\bf r})]^2r $ and for different basis functions
$Q_{AB} = \int d^3r \chi_A({\bf r}) \chi_B({\bf r})$. Since the
product of two gaussians with exponent $a$ centered on two neighboring
atoms a distance $d$ apart is a gaussian centered on the midpoint of the
two atoms with factor $e^{-ad^2/2}$ and there are three nearest neighbors,
then we have, approximately, $Q_{AB} \approx 3 e^{-ad^2/2} Q_{AA}$. 

Since the dominant exponent for our calculations is 0.2 (a.u.)$^{-2}$, the
ratio is $3 e^{-ad^2/2} = 1.4 > 1.0$. The value of 0.2 (a.u.)$^{-2}$
for the exponent $a$ does not depend on the chirality and size of the tube since it is a characteristic only of the local, spatial character of the bonds.
Therefore, it should be generally true that 
the term $\Delta_{AB}$ is dominant over $\Delta_{AA}$ and, consequently, 
$\Delta$ is negative. We expect the
dark state to be generally lower than the bright state for carbon nanotubes,
as we see in our {\it ab initio} calculations for several cases.
This is so, because the state ${\bf v}_+$, which has the higher
eigenvalue, $E_+$, is an antisymmetric combination of the transitions
at $K$ and $K'$ (the relative sign
is given by the factor $\Delta^*/|\Delta|$). 
  
For sufficiently long radiative lifetimes, exciton relaxation
to triplet states may also become relevant~\cite{pere+05nl}.  
We therefore briefly discuss the splitting between singlet and 
dipole-forbidden spin-triplet states. 

To calculate such splitting, we perform two BS calculations
with two different hamiltonians, one for the singlet state, $H_s = D + 2X - W$,
and one for the triplet state, given by $H_t = D - W$. These hamiltonians differ
only in the exchange term. Therefore, unlike the dark-bright
splitting of the singlet state, which is due to the {\it direct} Coulomb
interaction $W$, the singlet-triplet splitting is due to the {\it indirect exchange} Coulomb interaction.
Our {\it ab initio} calculations show that for the triplet states, 
the even exciton lies above the odd
one [i.e., we have $E^T_{1g}>E^T_{1u}$], 
which is a trend opposite to that found for the singlet states.
This behavior has been found also for zig-zag tubes~\cite{spat+05prl}. 
The above demonstration for singlet states
does not apply to the triplet states.
This is so because we have found numerically that the simple 
four-band picture (two pairs of conduction-valence bands) does not hold.
In fact, while for the singlet state $99\%$ of the weight is concentrated
on the $K$ and $K'$, for the triplet states at least $6\%$ of the weight
is concentrated elsewhere.
By evaluating $\Delta$ due to these other contributions, we find
a positive contribution which is enough to reverse the sign of $\Delta$,
forcing the even triplet exciton to lie above the odd one.

\begin{table}
\begin{ruledtabular}
\begin{tabular}{c c c c c}
$(n, m)$ & $d$ ($\AA$) &$E_{1u}^S-E_{1g}^S$ & $E_{1g}^S-E_{1g}^T$ &
$E_{1u}^S-E_{1u}^T$ \\\hline
(4,2) & 4.14 & 0.090 & 0.035 & 0.345   \\
(6,4) & 6.83 & 0.040 & 0.010 & 0.130   \\
(8,4) & 8.29 & 0.025 & 0.005 & 0.075   \\
\end{tabular}
\end{ruledtabular}
\caption{Energy splittings in eV. The third, fourth, and fifth columns are
the splittings for the $u-g$ singlet states, singlet-triplet for the gerade 
states and the single-triplet splitting for the ungerade states, respectively.}
\label{table}
\end{table}

The above symmetry analysis of the singlet states
suggests new possibilities for
increasing the quantum efficiency through symmetry breaking.
It can be deduced from the above discussion and perturbation theory
that the dark state becomes optically allowed, 
if the interaction which breaks the symmetry is of
the order of magnitude of the splitting $E_u-E_g \approx 0.1$~eV. 
One way to break the symmetry involves applying external fields.
However, their required magnitude may be too large for realistic 
technological applications, as shown in Ref.~\cite{zari+06prl},
where magnetic fields are used. 
An alternative strategy could involve the use of interacting semiconductor
nanotubes. This is suggested by recent theoretical work
on excitons in polimeric crystals, where interactions between different 
polimeric chains were shown to control the splitting between 
dark and bright excitons~\cite{ruin+02prl,buss+02apl}.
We have therefore reason to believe that 
bundles or arrays of {\it semiconducting} nanotubes could be examples 
of systems with increased quantum efficiency and possible candidates
of interest for practical devices. 

In conclusion, we have provided theoretical and numerical proof, using
an {\it ab initio}, many-body approach, and separately, a four-band model,
that dark excitonic states in semiconducting single-walled nanotubes exist, 
that they have even ${\bf R_U}$ symmetry,
and that they are always below the corresponding bright state. Furthermore,
the dark-bright energy for tubes in the diameter range of 0.4 - 0.8~nm is
some tens of meV with decreasing value as the tube radius increases. We show
that these results come from a four-band picture and that the splitting
is due to the direct Coulomb interaction between the electron-hole  and
the matrix element of the inter-valley interaction.
We also calculate the singlet-triplet splitting.
Here we find results of the same order of magnitude and the same 
decreasing trend with increasing radius as in the case of the 
dark-bright splitting. 
Finally, we have proposed, on the basis of the symmetry characteristics 
of the dark state, coupled semiconductor nanotubes as systems where 
the symmetry breaking induced by the intertube interaction leads to
increased quantum efficiency.

We thank C. Lienau and J. Kono for useful discussions. 
Computer time was partly provided
by CINECA through INFM Parallel Computing Projects.
The support by the RTN EU Contract ``EXCITING'' No.  HPRN-CT-2002-00317,
and by FIRB ``NOMADE'' is also acknowledged.

\end{document}